\documentclass[twocolumn,aps,prl,showpacs]{revtex4}

\usepackage{epsfig}

\begin{document}

\title {Quantum Imaging and Selection Rules in Triangular Quantum Corrals}
\author{Nikolaos A. Stavropoulos and Dirk K.~Morr}
\affiliation{Department of Physics, University of Illinois at
Chicago, Chicago, IL 60607}
\date{\today}
\begin{abstract}
We study quantum imaging in a triangular quantum corral that is
embedded in a superconducting host system with $s$-wave symmetry.
We show that the corral acts as a {\it quantum copying machine} by
creating multiple images of a {\it quantum candle}. We obtain new
selection rules for the formation of quantum images that arise
from the interplay of the corral's geometry and the location of
quantum candles. In more complex corral structures, we show that
quantum images can be projected ``around the corner".
\end{abstract}


\maketitle

Over the last few years, a growing number of exciting quantum
phenomena has been observed \cite{Man00,exp,Bra02,Pie04,Repp04}
that arise from the interplay between the geometry and quantum
properties of nanoscale atomic structures and their coupling to a
fermionic quantum many-body systems. Among these phenomena are the
formation of {\it quantum images} (also referred to as {\it
quantum mirages}) in elliptical quantum corrals \cite{Man00},
electronic lifetime effects in triangular quantum corrals
\cite{Bra02}, and magnetization effects in triangular Co islands
\cite{Pie04}. While the investigation of these phenomena is of
general fundamental interest, it could potentially lead to
important applications in the field of spin electronics and
quantum information technology~\cite{spinQC}. Theoretically, much
progress in understanding the interaction between nanostructures
and quantum many-body systems has been made by studying the
formation of a {\it single} quantum image in quantum corrals that
are embedded in metallic \cite{theory,review} or superconducting
(SC) \cite{Morr04} host systems.

In this Letter, we argue that complex nanoscale structures are
prototype systems for the observation of novel quantum phenomena.
In particular, we demonstrate that a triangular quantum corral can
be used as a {\it quantum copying machine} that creates multiple
quantum images of characteristic features (so-called {\it quantum
candles}) in the host system's density-of-states (DOS). As a
quantum candle, we employ the spectroscopic signature of a
fermionic bound state induced by a magnetic impurity in a
superconducting host system with $s$-wave symmetry. We show that
the formation of quantum images inside a triangular corral
consisting of {\it non-magnetic} impurities is determined by a set
of selection rules that arises from the interplay between the
corral's geometry and the location of quantum candles. Moreover,
we demonstrate that such a corral can suppress the formation of
fermionic bound states, leading to the important result that {\it
non-magnetic impurities} can reverse the pair-breaking effect of a
{\it magnetic defect}. Finally, we show that {\it double
triangular corrals} allow the projection of quantum images
``around the corner", opening the interesting possibility to
custom design the imaging properties of quantum corrals.

In order to study novel quantum effects arising from the
interaction of a triangular quantum corral with a superconducting
host system, we employ a generalized scattering $\hat{T}$-matrix
theory \cite{Morr04,Morr03a,Shiba68}. The host system's local
Greens function (in Nambu-notation) in the presence of the corral
is given by
\begin{eqnarray}
\hat{G}({\bf r},{\bf r'},\omega_n)&=&\hat{G}_0({\bf r},{\bf r'},\omega_n) \nonumber \\
& & \hspace{-2cm} +\sum_{i,j=1}^N \hat{G}_0({\bf r},{\bf
r}_i,\omega_n)\hat{T}({\bf r}_i,{\bf r}_j,\omega_n)\hat{G}_0({\bf
r}_j,{\bf r'},\omega_n) \ , \label{Ghat}
\end{eqnarray}
where the sum runs over the locations ${\bf r}_i \ (i=1,..,N)$ of
the $N$ impurities forming the corral. The ${\hat T}$-matrix
follows from the Bethe-Salpeter equation
\begin{eqnarray}
\hat{T}({\bf r}_i,{\bf r}_j,\omega_n)&=&\hat{V}_i \, \delta_{i,j}
\nonumber \\
& & \hspace{-1.5cm} +\hat{V}_i \, \sum_{l=1}^N \hat{G}_0({\bf
r}_i,{\bf r}_l,\omega_n)\hat{T}({\bf r}_l,{\bf r}_j,\omega_n) \ ; \\
\hat{V}_i&=&\frac{1}{2} \left(U_i \sigma_0 + J_iS \sigma_3
\right)\tau_3 \ , \label{Tmatrix}
\end{eqnarray}
and the electronic Greens function of the unperturbed (clean)
system in momentum space is
\begin{equation}
\hat{G}^{-1}_0({\bf k},i\omega_n)=\left[ i\omega_n \tau_0 -
\epsilon_{\bf k} \tau_3 \right] \sigma_0 + \Delta_0 \tau_2
\sigma_2  \ . \label{G0}
\end{equation}
$U_i (J_i)$ is the potential (magnetic) scattering strength of the
impurity at site ${\bf r}_i$, $S$ is the spin of a magnetic
impurity, and ${\bf \sigma}$, ${\bf \tau}$ are the Pauli-matrices
in spin and Nambu space, respectively. $\Delta_{0}$ is the
superconducting $s$-wave gap, and $\epsilon_{\bf k}$ is the host
system's normal state dispersion. In this approach, magnetic
impurities are treated as classical variables \cite{Shiba68}
(corresponding to a large-$S$ limit) since $J$ is taken to be
smaller than the critical value $J_c$ necessary for a Kondo-effect
to occur in an s-wave superconductor \cite{Sat92}, in full
agreement with experiment \cite{Yaz97}. Finally, the local DOS,
$N({\bf r},\omega)$, shown below, is obtained numerically from
Eqs.(\ref{Ghat})-(\ref{G0}) with $N({\bf
r},\omega)=A_{11}+A_{22}$, $A_{ii}({\bf r},\omega)=-{\rm Im}\,
\hat{G}_{ii}({\bf r},\omega+i\delta)/ \pi$ and $\delta=0.1$ meV.

\begin{figure}[t]
\epsfig{file=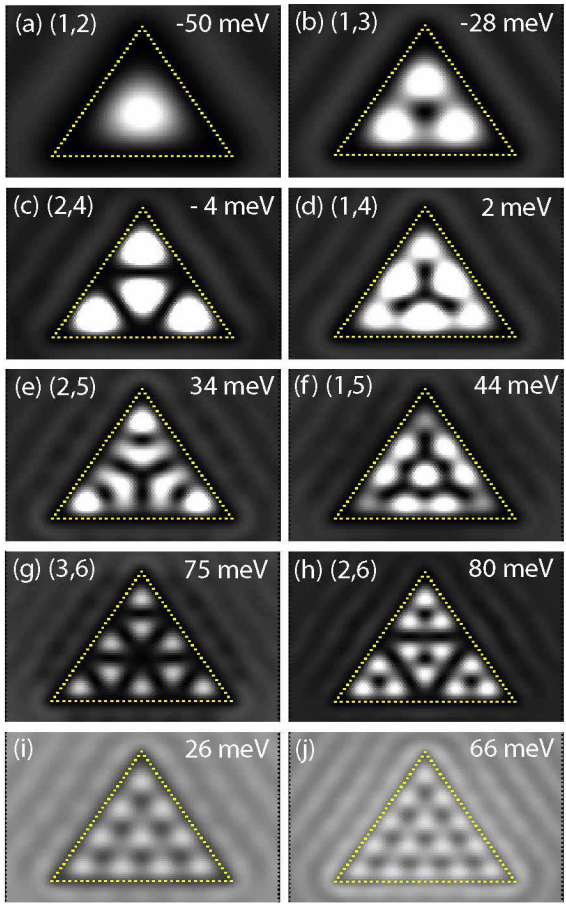,width=7.5cm} \caption{(a)-(h) Spatial DOS
plot for the eight lowest energy eigenmodes (eigenmode energy is
shown in the upper right corner) in the unitary scattering limit
($U_i=4$ eV). (m,l) are the quantum numbers of the corresponding TPW
eigenstate. (i),(j) Eigenmodes of the corral for $U_i=0.5$ eV.}
\label{DOSTriangle}
\end{figure}

Motivated by recent experiments \cite{Bra02}, we first study the
eigenmode spectrum of a triangular equilateral quantum corral
embedded in a normal host system. To facilitate comparison with
experiment, we consider a two-dimensional host system with a
triangular lattice (lattice constant $a_0=1$) and $\epsilon_{\bf
k}= k^2/2m-\mu$ ($\hbar=1$) where $\mu=-65$ meV is the chemical
potential and $k_F=0.24$ the Fermi wave-vector (qualitatively
similar results to the ones shown below are expected for a 3D host
system). In Fig.~\ref{DOSTriangle}(a)-(h) we present a spatial DOS
plot of the eight lowest energy eigenmodes [with light (dark)
regions indicating a large (small) DOS] for a corral consisting of
90 {\it non-magnetic} impurities with $U_i=4$ eV (the impurities
are represented by filled yellow squares, separated by $\Delta
r=2$). In this unitary scattering limit, the eigenmodes (i.e.,
their spatial structure and ordering in energy) are well described
by the eigenstates, $\phi_{lm}$, of an infinitely deep triangular
potential well (TPW) \cite{Kri82} [the corresponding quantum
numbers $(m,l)$ are shown in the upper left corner of
Figs.~\ref{DOSTriangle}(a)-(h)].
%
%
With decreasing $U_i$, the energy separation of the eigenmodes is
reduced. In addition, new eigenmodes emerge, such as the ones
shown in Figs.~\ref{DOSTriangle}(i),(j) for $U_i=0.5$ eV, which
are similar to those observed experimentally (cf. Fig.1(d) in
Ref.~\cite{Bra02}).

The spatial form of the corral's eigenmodes opens the possibility
to form multiple {\it quantum mirages}. To demonstrate the
qualitative nature of this effect, we retain the above parameters
and take the superconducting gap to be $\Delta_0=4$ meV, yielding
a superconducting coherence length $\xi_c=k_F/(m\Delta_0)=135$. As
the quantum candle whose image is formed we use the spectroscopic
signature of a fermionic bound state induced by a single magnetic
impurity ($J_i=1.0$ eV), located in the center of the corral at
${\bf r}_1=(0,0)$.
%
%
\begin{figure}[t]
\epsfig{file=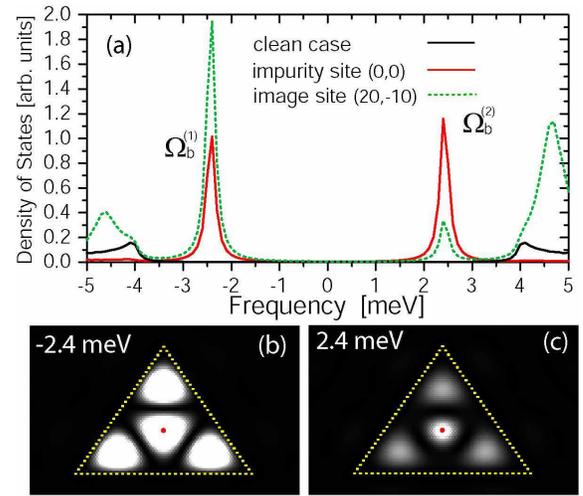,width=7.5cm} \caption{(a) DOS as a function
of frequency. (b),(c) Spatial DOS plot at the bound state energies
$\Omega_b^{(1,2)}=\mp 2.4$ meV. } \label{DOSTriangle1imp}
\end{figure}
This signature consists of a particlelike and holelike peak in the
DOS at energies $\Omega_b^{(1,2)} = \mp 2.4$ meV, as shown in
Fig.~\ref{DOSTriangle1imp}a. In Figs.~\ref{DOSTriangle1imp}(b),(c)
we present spatial DOS plots at $\Omega_b^{(1,2)}$ (the location
of the magnetic impurity is shown as a filled red circle). The
formation of the impurity bound state is accompanied by the
excitation of the $(2,4)$-eigenmode [Fig.~\ref{DOSTriangle}(c)]
and by the emergence of three images of the bound state peaks
inside the corral. Note that only eigenmodes that possesses
sufficiently large spectral weight at the impurity site and are
close in energy to $\Omega_b^{(1,2)}$ are relevant for the
formation of quantum images. Since the energy of the
$(2,4)$-eigenmode, $E_{(2,4)}=-4$ meV, is closer to
$\Omega_b^{(1)}$ than to $\Omega_b^{(2)}$, the spectral weight of
the quantum images is larger at $\Omega_b^{(1)}$ than at
$\Omega_b^{(2)}$. This result demonstrates that a triangular
quantum corral acts as a {\it quantum copying machine} for
distinct features in the DOS. Moreover, the corral's imaging
properties can be specifically designed since changing the
corral's size leads to a shift in the eigenmode energies
\cite{Kri82}. For example, in a corral consisting of 117
impurities, the imaging properties in the SC state are determined
by the $(1,5)$-mode [$E_{(1,5)}=0$ meV] leading to a different
spatial pattern of the quantum images \cite{MorrFP}. Finally, we
note as an important result that the formation of an impurity
bound state can be completely suppressed inside the corral. To
demonstrate this effect, we place the magnetic impurity at a node
[${\bf r}_1=(-5,-5)$] of the $(2,4)$- and $(1,4)$-eigenmodes, as
shown in Fig.~\ref{DOSSuppression}(a) for $|\omega|>\Delta_0$. In
this case, the DOS at ${\bf r}_1$ [Fig.~\ref{DOSSuppression}(b)]
does not possess any signature of an induced bound state and
hence, no image is observed anywhere inside the corral. This
complete suppression arises from the incompatibility of the bound
state with the boundary conditions provided by the corral's wall.
In other words, an impurity bound state can only be formed if it
can couple to one of the corral's eigenmodes. The importance of
this result lies in the fact that while non-magnetic impurities
cannot induce a fermionic bound state in an $s$-wave
superconductor, they can suppress its formation and thus reverse
the pair-breaking effect of a {\it magnetic defect}.
%
%
\begin{figure}[h]
\epsfig{file=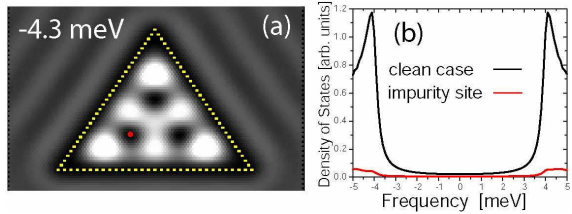,width=7.5cm} \caption{(a) Spatial DOS plot
for $|\omega|> \Delta_0$. (b) DOS at the site of the magnetic
impurity.} \label{DOSSuppression}
\end{figure}

\begin{figure}[t]
\epsfig{file=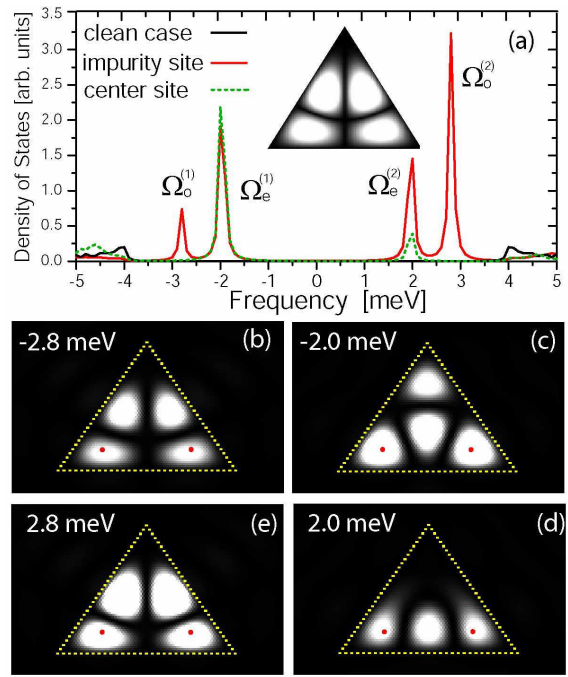,width=7.5cm} \caption{(a) DOS as a function
of frequency. Inset: Spatial plot of $|\phi^{(o)}_{14}|^2$. (b)-(e)
Spatial DOS plot for the odd [(b) and (e)] and even [(c) and (d)]
impurity bound states.} \label{DOSTriangle2imp}
\end{figure}

Next, we consider the effects of two magnetic impurities located
inside the corral at ${\bf r}_1=(-10,-10)$ and ${\bf
r}_2=(20,-10)$. The angle, $\alpha$, between the impurity spins is
determined by the interaction between them. In what follows, we
assume a ferromagnetic alignment of the spins ($\alpha=0$),
however, results similar to those shown below are also obtained
for $\alpha \not = 0$. Quantum interference of scattered
electronic waves leads to the formation of even and odd impurity
bound states (with respect to a vertical axis midway between the
two impurities) and a splitting of the bound state energies
\cite{com1}. As a result, the DOS exhibits four peaks at
$\Omega_o^{(1,2)}=\mp 2.8$ meV and $\Omega_e^{(1,2)}=\mp 2.0$ meV
[see Fig.~\ref{DOSTriangle2imp}(a)]. The formation of quantum
images requires that the even/odd bound states couple to corral
eigenmodes of the same symmetry. The eigenstates of a TPW and thus
the corral eigenmodes transform under reflection at the vertical
axis as  $\phi_{lm} \rightarrow -\exp\left[ i 2 \pi (m+l)/3
\right] \phi^*_{lm}$ \cite{Kri82}. While $\phi_{24}$ is imaginary
and thus even under reflection, an even and odd wave-function is
formed from $\phi_{14}$ via $\phi^{(e,o)}_{14}(x)=\phi_{14}(x) \pm
\phi_{14}(-x)$.
%
%
In Figs.~\ref{DOSTriangle2imp}(b),(e) we plot the DOS at
$\Omega_{o}^{(1,2)}$ whose spatial form agrees well with that of
$|\phi^{(o)}_{14}|^2$ shown in the inset of
Fig.~\ref{DOSTriangle2imp}(a). We therefore conclude that
$\Omega_{o}^{(1,2)}$ are the frequencies of the odd bound state,
while $\Omega_{e}^{(1,2)}$ are the energies of the even bound
state whose spatial DOS is shown in
Figs.~\ref{DOSTriangle2imp}(c),(d). The imaging properties of the
corral are thus frequency dependent due to the interplay between
the corral's geometry and the location of the quantum candles.

\begin{figure}[t]
\epsfig{file=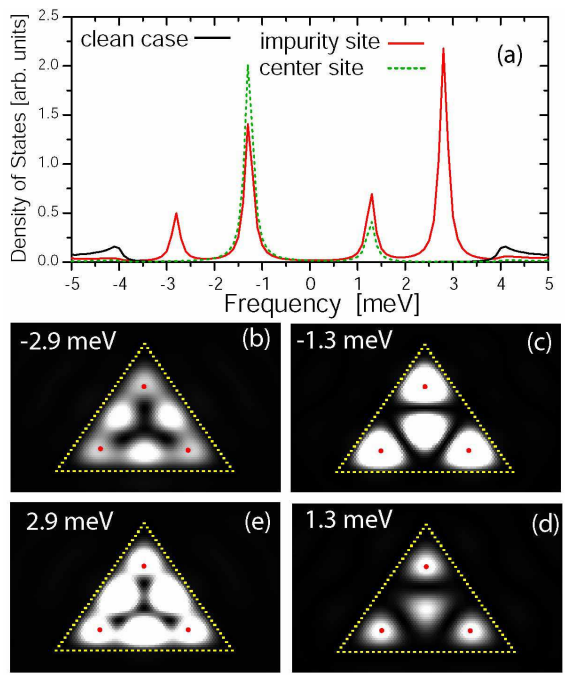,width=7.5cm} \caption{(a) DOS as a function
of frequency. (b)-(d) Spatial plot of the DOS for the impurity bound
state with $n=2$ [(b) and (e)] and $n=0$ [(c) and (d)]. }
\label{DOSTriangle3imp}
\end{figure}

This interplay can be further studied by placing three magnetic
impurities with parallel spins at the corners of an equilateral
triangle at ${\bf r}_1=(-10,-10)$, ${\bf r}_2=(20,-10)$, and ${\bf
r}_3=(-10,20)$. Since the degeneracy of the impurity bound states
is again lifted via quantum interference, we expect to find six
peaks in the DOS. Instead, the DOS exhibits four peaks, as shown
in Fig.~\ref{DOSTriangle3imp}(a), corresponding to the presence of
only two non-degenerate impurity states.
%
%
This reduction to two impurity states arises from a new type of
selection rule that is based on the interplay between the corral
geometry and the location of the quantum candles. Under a rotation
of $2\pi/3$ around the corral's center, $\phi_{lm}$, and hence the
corral eigenmodes, transform as $\phi_{lm} \rightarrow \exp [ i 2
\pi (m+l)/3  ] \phi_{lm}$. Due to their geometry the
non-degenerate impurity bound states possess the same
transformation properties, and their formation thus requires that
they couple to eigenmodes with $n=(m+l){\rm mod} 3=1,2,3$.
However, the eigenmodes with $n=1$ are at energies $|E_{(m,l)}|
\gg \Delta_0$, thus preventing the creation of the bound state
with $n=1$. As a result, only the bound states with $n=0$
[Figs.~\ref{DOSTriangle3imp}(c) and (d)] and $n=2$
[Figs.~\ref{DOSTriangle3imp}(b) and (e)] are formed via their
coupling to the $(2,4)$- [Fig.~\ref{DOSTriangle}(c)] and
$(1,4)$-eigenmodes [Fig.~\ref{DOSTriangle}(d)], respectively.

\begin{figure}[t]
\epsfig{file=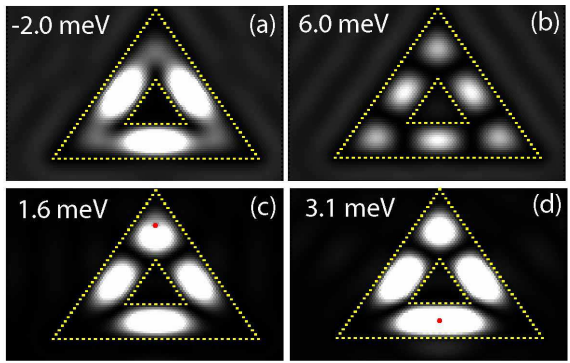,width=7.5cm} \caption{Normal state eigenmodes
of a double triangular quantum corral. (c),(d) Spatial DOS plot for
two different locations of a magnetic impurity with (c)
$\Omega_b^{(2)}=1.6$ meV and (d) $\Omega_b^{(2)}=3.1$ meV.}
\label{DoubleTriangle}
\end{figure}

More complex corral structures can be employed to project quantum
images ``around the corner". To demonstrate this effect, we insert
a triangular corral with 42 {\it non-magnetic} impurities ($U_i=4$
eV) into the corral discussed above. In the normal state, two
eigenmodes of this double corral [Figs.~\ref{DoubleTriangle}(a)
and (b)] possess energies that render them relevant for the
formation of quantum images.
%
%
By placing a magnetic impurity ($J_i=2$ eV) between the apices of
the triangles in the SC state [Fig.~\ref{DoubleTriangle}(c)] a
quantum image is formed ``around the corner" between the
triangles' bases. Similarly, a quantum image is created between
the apices by placing the magnetic impurity between the triangles'
bases [Fig.~\ref{DoubleTriangle}(d)]. Note the significant shift
in $\Omega_b$ when the position of the impurity is changed. This
result opens the interesting possibility to custom design the
imaging properties of quantum corrals to form mirages at arbitrary
locations.

Finally, we note that the qualitative features of our results
presented above are robust against changes in the band parameters
or $\Delta_0$, as long as $\xi_c$ is larger than the size of the
corral. Moreover, whether a Kondo-effect can occur inside a corral
and how it is affected by the corral's eigenmodes is an
interesting but non-trivial question whose study we reserve for
future work \cite{MorrFP}.

In summary, we demonstrate that the eigenmode spectrum of a
triangular quantum corral can be employed (i) to create multiple
images of a quantum candle, and (ii) to suppress the formation of
impurity bound states. We obtain new selection rules for quantum
imaging that arise from the interplay of the corral's geometry and
the location of quantum candles. Finally, we show that more
complex nanostructures allow the projection of quantum images
``around the corner".

We would like to thank K.-F. Braun, J.C. Davis, O. Pietzsch, and
R. Wiesendanger for stimulating discussions, and would especially
like to thank K.-H. Rieder for a series of discussions that
motivated this work. D.K.M. acknowledges financial support from
the Alexander von Humboldt foundation.


\begin{thebibliography}{99}
\bibitem{com1} No even/odd bound states are formed for $\alpha =
\pi$ \cite{Morr04}.
\end{thebibliography}
\end{document}